\documentstyle[12pt,epsfig]{article}
\def\b{\bar}
\def\d{\partial}

\def\cA{{\cal A}}
\def\cD{{\cal D}}

\def\M{\mathcal{M}}
\def\m{\mu}
\def\n{\nu}

\def\t{\tau}

\def\~{\tilde}
\def\h{\eta}

\def\bY3{\bar Y_{,3}}
\def\Y3{Y_{,3}}
\def\z{\zeta}
\def\Z{{\b\zeta}}
\def\Y{{\bar Y}}
\def\cZ{{\bar Z}}
\def\`{\dot}
\def\be{\begin{equation}}
\def\ee{\end{equation}}
\def\bea{\begin{eqnarray}}
\def\eea{\end{eqnarray}}

\def\fn{\footnote}

\def\cF{{\cal F}}

\def\mn{{\mu\nu}}

\begin{document}

\title{Axial Stringy System of the Kerr Spinning Particle
\fn{Paper in honor of the 100th anniversary of Dmitrii Ivanenko birthday.}}

\author{Alexander Burinskii\\
Gravity Research Group, NSI Russian
Academy of Sciences\\
B. Tulskaya 52, 115191 Moscow, Russia}
\maketitle

\begin{abstract}
\noindent
The structure of classical spinning particle based on the
Kerr-Newman black hole (BH) solution  is investigated.
For  large angular momentum, $|a|>>m$, the BH horizons
disappear exposing a naked ringlike source which is  a
circular relativistic  string. It was shown recently that
electromagnetic excitations of this string lead to the appearance
of an extra axial stringy system which consists of two half-infinite
strings of opposite chirality.
In this paper we consider the relation of this stringy system
to the Dirac equation.

We show that the axial strings are
the Witten superconducting strings and describe their structure
by the Higgs field model
where the Higgs condensate is used to regularize  axial
singularity. We argue that this axial stringy system may play the
role of a classical carrier of the wave function.

\end{abstract}

\section{Introduction}
The Kerr rotating black hole solution displays some remarkable
relations to spinning particles
\cite{Car,Isr,Bur0,IvBur,Lop,BurSen,BurStr,BurSup,BurBag}.
For the parameters of elementary particles, $|a|>> m$, and black-hole
horizons disappear. This  changes drastically the usual black hole
image since there appear  the rotating source
in the form of a closed singular ring of the Compton radius $ a=J/m$.
\fn{Here $J$ is angular momentum and $m$ is mass. We use the units $c=\hbar
=G=1$, and signature $(-+++)$.}.
In this work we continue investigation of
the classical spinning particle based on the
Kerr-Newman black hole (BH) solution.

One of the first papers of this
series was the work \cite{Bur0} published 30 years ago.
In this paper the model of the Kerr
spinning particle - ``microgeon" was suggested which was based on the
Wheeler's idea of the ``mass without mass''.
The Kerr ring was considered as a
gravitational waveguide for the traveling electromagnetic (and
fermionic) wave excitations.

That time the central Russian
theoretical seminars in Moscow were hold by Dmitrii Dmitrievich Ivanenko at
Moscow State University.  At this seminar I learnt what is the Kerr
solution and strings.
My first talk on the geon model, about 1972, was met by
Dmitrii Dmitrievich with some degree of skepticism.
However later, after discussions, his opinion was changed,
and in 1975 we published a common letter on ``Gravitational strings
in the models of elementary particles" \cite{IvBur} which contained the
conjecture that the Kerr singular ring was the string.
By preparing of this paper Dmitrii Dmitrievich said
``It may be very important if it is only true in a small degree."

The assumption that the Kerr singular ring is the string was based on
some evidences of Refs. \cite{BurSen,IvBur1,Bur1}. However, the attempts
to show it rigorous ran into obstacles which were related with
the very specific motion of the Kerr ring - the lightlike sliding
along itself.  It could be described as a string containing
lightlike modes of only one direction.  However, the system of the bosonic
string equations does not admit such solutions.
Only thirty years later  this problem
was resolved. In the recent paper \cite{BurOri}, it was shown
that the Kerr ring satisfies all the stringy equations representing
a string with an orientifold world sheet.

In this paper we discuss in details the second stringy structure of
the Kerr spinning particle which was obtained only recently \cite{BurTwo}.
\fn{It should be noted, that in our old common paper with Dmitrii Dmitrievich
\cite{IvBur} the existence of this second stringy system
in the Kerr spinning particle was mentioned too.}
We show that the aligned electromagnetic
excitations of the Kerr circular string
unavoidably lead to the appearance of the axial half-infinite strings
which are  similar to the Dirac monopole string and
topologically coupled to the Kerr ring.
The class of the aligned e.m. solutions turns out to be very restricted:
all the solutions can be numbered by the integer index
$n=0, \pm 1, \pm 2 ...$, and, for the exclusion $n=0$, they lead to the
appearance of axial half-infinite singularities.  We show that
these axial strings  carry the chiral
traveling waves induced by the e.m. excitations of the Kerr circular string.

\begin{figure}[ht]
\centerline{\epsfig{figure=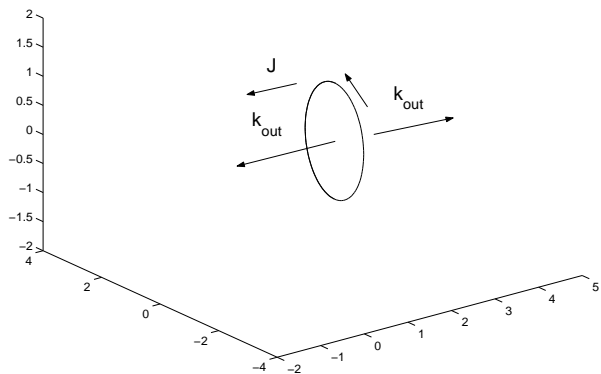,height=6cm,width=8cm}}
\caption{Stringy skeleton of the Kerr spinning particle.
Circular D-string and the directed outwards two axial half-infinite chiral
D-strings.}
\end{figure}

Therefore, the frame of the Kerr spinning particle turns out to be
consisting of two topologically coupled stringy systems.
The appearance of the
axial half-infinite strings looks strange at first sight. However, we
obtain that it can be a new and very important element of
the structure of spinning particles, since it has
the relation  to the Dirac equation. We
show that for the moving particle the excitations of the chiral strings
are modulated by de Broglie periodicity, and therefore, the axial stringy
system turns out to be a carrier of the wave function.

In the zone which is close to the Kerr string, our treatment is based on
the Kerr-Schild formalism  \cite{DKS}
and previous paper \cite{Bur-nst} where the real and complex
structures of the Kerr geometry were considered. For the reader convenience
 we describe briefly the necessary details of these structures.
Meanwhile, in the far zone, structure of this string is
described by the very simple class of pp-wave solutions \cite{KraSte,Per}.
The resulting stringy frame turns out to be very simple and easy
for description. We obtain that these strings belong to the class of
the chiral superconducting strings
which have recently paid considerable attention in astrophysics
\cite{CarPet,Vil,BOV}.
In fact these chiral strings turns out to be the Witten's superconducting
strings \cite{Wit,VS}. Note, that the similar chiral strings are also very
popular in different models of the high dimensional superstring theory,
forming the fundamental strings \cite{Dab}, chiral null systems
\cite{HorTse}, multiply wound strings \cite{LunMat}
and supertubes \cite{MatTow}.

During the treatment we meet some singular
structures and divergences  which  have to be regularized. In particular, the mass of the infinite
axial stringy system is divergent and we perform its renormalization,
which shows that tension of the axial stringy system tends to zero
for a free particle, but can take a finite value for a bounded system.
The field singularities of the stringy pp-waves have to be also regularized,
which is realized by introduction of a superconducting source for
these strings.
It is remarkable that the considered procedures of regularization,
having the clear physical meaning in the stringy Kerr geometry,
resemble the regularizations used in QED \cite{Sch,AhiBer}.

\section{The  Kerr geometry and the Kerr  circular string.}

We use the Kerr-Schild approach to the Kerr geometry \cite{DKS},
which is based on the Kerr-Schild form of the metric \be g_{\m\n}
= \h_{\m\n} + 2 h k_{\m} k_{\n}, \label{ksa} \ee where $ \h_\mn $
is the metric of auxiliary Minkowski space-time, $ h= \frac
{mr-e^2/2} {r^2 + a^2 \cos^2 \theta},$ and $k_\m$ is a twisting
null field, which is tangent to the Kerr principal null congruence
(PNC) and is determined by the form \fn{The rays of the Kerr PNC
are twistors and the Kerr PNC is determined by the Kerr theorem as
a quadric in projective twistor space \cite{Bur-nst}.}
 \be k_\m dx^\m = dt +\frac z r dz + \frac r {r^2 +a^2} (xdx+ydy) - \frac a
{r^2 +a^2} (xdy-ydx) .  \label{km} \ee The form of the Kerr PNC is
shown in Fig. 2.  It follows from Eq.(\ref{ksa}) that the field
$k^\m$ is null with respect to $\h_\mn$ as well as with respect to
the full metric $g_\mn$, \be k^\m k_\m = k^\m k^\n g_\mn =  k^\m
k^\n \eta_\mn. \label{kgh} \ee

\begin{figure}[ht]
\centerline{\epsfig{figure=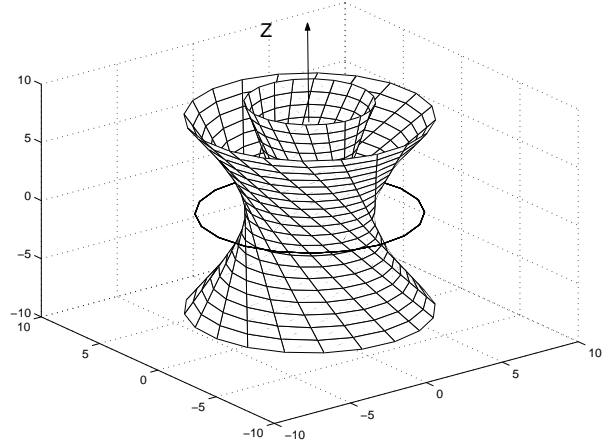,height=6cm,width=8cm}}
\caption{The Kerr singular ring and 3D section of the Kerr
principal null congruence. Singular ring is a branch line of
space, and PNC propagates from the ``negative'' sheet of the Kerr
space to the ``positive '' one, covering the space-time twice. }
\end{figure}

The metric is singular at the ring $r=\cos\theta=0$, which is
the focal region of the oblate spheroidal coordinate system
$r, \theta, \phi$.

  The Kerr singular ring is the branch line of the
Kerr space on two folds: positive sheet ($r>0$) and `negative' one
($r<0$). Since for $|a|>>m$ the horizons disappear, there appears
the problem of the source of the Kerr solution with the
alternative: either to remove this twofoldedness or to give it a
physical interpretation.  Both approaches have received attention,
and it seems that both are valid for different models. The most
popular approach was connected with the truncation of the negative
sheet of the Kerr space, which leads to the source in the form of
a relativistically rotating disk \cite{Isr} and to the class of
the disklike \cite{Lop} or baglike \cite{BurBag} models of the Kerr
spinning particle.

 An alternative way is to retain the negative sheet treating it as
 the sheet of advanced fields. In this case the source of the spinning
particle turns out to be the Kerr singular ring with the
electromagnetic excitations in the form of traveling waves, which
generate spin and mass of the particle.  A model of this sort was
suggested in 1974 as a model of ``microgeon with spin"
\cite{Bur0}. The Kerr singular ring was considered as a
 waveguide providing a circular propagation of an electromagnetic or
fermionic wave excitation. Twofoldedness of the Kerr geometry
admits the integer and half integer excitations  with $n=2\pi
a/\lambda$ wave periods on the Kerr ring of radius $a$, which
turns out to be consistent with the corresponding values of the
Kerr parameters $m= J/a$.

The lightlike structure of the Kerr ring worldsheet is seen from
the analysis of the Kerr null congruence near the ring. The
lightlike rays of the Kerr PNC are tangent to the ring.

It was recognized long ago \cite{IvBur} that the Kerr singular ring
 can be considered in the Kerr spinning particle as a string with
traveling waves.  One of the most convincing evidences obtained
by the analysis of the axidilatonic generalization of the Kerr
solution (given by Sen \cite{Sen}) near the Kerr singular ring
was given in \cite{BurSen}. It was shown that the fields
 near the Kerr ring are very similar to the field around a
heterotic string.

The analysis given in \cite{BurOri} showed
that the worldsheet of the Kerr ring satisfies the
bosonic string equations and constraints; however, there appear
problems with boundary conditions, and
these difficulties can be removed by the formation of the
worldsheet orientifold.

The interval of an open string  $\sigma
\in [0,\pi] $ has to be extended to $[0,2\pi]$, setting
\be X_R (\sigma
+\pi) =X_L (\sigma), \qquad X_L (\sigma +\pi) =X_R (\sigma).  \label{ext} \ee

By such an extension, there appear the both types of modes, ``right" and
``left"  since the ``left" modes play
the role of ``right" ones on the extended piece of interval. If
the extension is completed by the changing of orientation on the
extended piece,  $\sigma ^\prime = \pi - \sigma $, with a
subsequent identification of $\sigma$ and  $\sigma ^\prime$, then
one obtains the closed string on the interval $[0,2\pi]$ which is
 folded and takes the form of the initial open string.

Formally, the worldsheet orientifold represents a doubling of the
worldsheet with the orientation reversal on the second sheet. The
fundamental domain $[0,\pi]$ is extended to $\Sigma=[0,2\pi]$ with
formation of folds at the ends of the interval $[0,\pi]$.

\section{Aligned e.m. solutions on the Kerr-Schild background}
\bigskip
To realize the idea of the Kerr spinning particle as a ``microgeon''
we have to consider the  electromagnetic excitations of the Kerr string
which are described on the Kerr background
by the aligned to the Kerr PNC solutions. Such e.m. solutions can be treated
as candidates for the self-consistent Einstein-Maxwell solutions.

The treatment on this section is based on the Kerr-Schild formalism,
and the readers which are not aware of this formalism can omit this part
by first reading going to the physical consequences of these solutions.

The aligned field equations for the Einstein-Maxwell system in the
Kerr-Schild class were obtained in \cite{DKS}.
Electromagnetic field is given by tetrad components of self-dual tensor
\be \cF _{12} =AZ^2 \label{1}\ee
\be \cF _{31} =\gamma Z - (AZ),_1  \ . \label{2}\ee
The equations for electromagnetic field are
\be A,_2 - 2 Z^{-1} \cZ Y,_3 A  = 0 , \label{3}\ee
\be \cD A+  \cZ ^{-1} \gamma ,_2 - Z^{-1} Y,_3 \gamma =0 .
\label{4}\ee
Gravitational field equations yield
\be M,_2 - 3 Z^{-1} \cZ Y,_3 M  = A\bar\gamma \cZ ,  \label{5}\ee
\be \cD M  = \frac 12 \gamma\bar\gamma  , \label{6}\ee
where
\be
\cD=\d _3 - Z^{-1} Y,_3 \d_1 - \cZ ^{-1} \Y ,_3 \d_2   \ . \label{cD}
\ee
Solutions of this system were given in \cite{DKS} only for stationary case
for $\gamma=0$.
Here we consider the oscillating electromagnetic
solutions which corresponds to the case $\gamma \ne 0$.

For the sake of simplicity we have to consider the gravitational
Kerr-Schild field as stationary, although in the resulting e.m.
solutions the axial symmetry is broken, which has to lead to
oscillating backgrounds if the back reaction is taken into account.

The recent progress in the obtaining the nonstationary solutions of the
Kerr-Schild class is connected with introduction of a complex retarded time
parameter $\t = t_0 +i\sigma = \t |_L$ which is determined as a result of the
intersection of the left (L) null plane
and the complex world line \cite{Bur-nst}. The left null planes are the
left generators of the complex null cones and play a role of the null cones
in the complex retarded-time construction.
The $\t$ parameter satisfies to the relations
\be
(\t),_2 =(\t),_4 = 0 \ . \label{7}
\ee
It allows one to represent the equation (\ref{3}) in the form
\be
(AP^2),_2=0 \ , \label{8}
\ee
and to get the following general solution
\be A= \psi(Y,\t)/P^2
\label{A}\ee
which has the form obtained in \cite{DKS}.
The only difference is in the extra dependence of the
function $\psi$ from the retarded-time parameter $\t$.

It was shown in \cite{Bur-nst} that action of operator $\cD$ on the variables
$Y, \bar Y $ and $ \rho$ is following
\be \cD Y = \cD \bar Y = 0,\qquad
\cD \rho =1 \ , \label{10}\ee
and therefore
$\cD \rho = \d \rho / \d t_0 \cD t_0  = P\cD t_0 =1 $,
that yields
\be
\cD t_0 = P^{-1} .
\ee
As a result the equation (\ref{4}) takes the form
\be
\dot A = -(\gamma P),_{\bar Y} , \label{11}
\ee
where $\dot {( \ )} \equiv \d_{t_0}$.

For considered here stationary background $P=2^{-1/2}(1+Y\bar Y)$, and
$\dot P = 0$.  The coordinates $Y$,  and $\t$ are independent from
$\bar Y$, which allows us to integrate Eq. (\ref{11}) and
we obtain the following general solution
\be
\gamma  = - P^{-1}\int \dot A d\bar Y =
 - P^{-1}\dot \psi (Y,\t) \int  P^{-2}d\bar Y  =
 \frac{2^{1/2}\dot \psi} {P^2 Y} +\phi (Y,\t)/P ,
\label{12}\ee
where $\phi$ is an arbitrary analytic function of $Y$ and $\t$.

The term $\gamma$  in
$ \cF _{31} =\gamma Z - (AZ),_1  \ $
describes a part of the null
electromagnetic radiation   which
falls of asymptotically as $1/r$ and
propagates along the Kerr principal null congruence $e^3$.
As it was discussed in \cite{Bur-nst,BurOri} it describes
a loss of mass by radiation with the stress-energy tensor
$\kappa T^{(\gamma)}_\mn = \frac 12 \gamma \bar \gamma e^3_{\m} e^3_{\n}$
and has to lead to an infrared divergence.
However,  the Kerr twofoldedness
and the structure of the Kerr principal null congruence show us that the
loss of mass on the positive sheet of metric is really compensated by an
opposite process on the ``negative" sheet of the Kerr space where is an
in-flow of the radiation. In the microgeon model
\cite{Bur-nst,BurOri,BurTwo}, this field acquires interpretation of the vacuum zero
point field $T^{(\gamma)}_\mn = <0|T_{\mn}|0>$.
Similar to the treatment of the zero point field in the
Casimir effect one has to regularize stress energy tensor
by the subtraction
\be T^{(reg)}_{\mn} = T_{\mn} - <0|T_{\mn}|0>, \ee
under the condition $ T^{(\gamma) \ \mn} ,_\m = 0$ which is
satisfied for the $\gamma$ term.

Let's now consider in details the second term in (\ref{2}):
\be
(AZ),_1 = (Z/P)^2 (\psi ,_Y - 2 \psi P_Y) +
(Z/P^2) \dot \psi \t,_1 + A Z,_1 .
\ee
For stationary case we have relations $Z,_1 =2ia \bar Y (Z/P)^3 $
and  $\t,_1 =- 2ia \bar Y Z/P^2 $ .
This yields
\be
(AZ),_1 = (Z/P)^2 (\psi ,_Y - 2ia  \dot \psi \bar Y /P^2 - 2 \psi P_Y/P) +
A 2ia \bar Y (Z/P)^3 .
\label{AZ1}
\ee

Since $Z/P =1/(r+ia \cos \theta)$, this expression contains the terms
which are singular at the Kerr ring and fall off like $r^{-2}$ and  $r^{-3}$.
However, it contains also
the factors which depend on coordinate
$Y = e^{i\phi} \tan \frac {\theta} 2 $ and
can be singular at the z-axis.

These singular factors  can be selected in the full
expression for the aligned e.m. fields and as a result there
appear two  half-infinite lines of
singularity,  $z^+$ and $z^-$,
which correspond to $\theta =0$ and $\theta=\pi$ and coincide
with corresponding axial lightlike rays of the Kerr principal null
congruence.  On the ``positive'' sheet of the Kerr background these two
half-rays are directed outward. However, one can see that they are going
from the ``negative" sheet and  appear on the ``positive'' sheet
passing through the Kerr ring (see Fig. 2).

The general solution for the aligned electromagnetic fields has the form

\be
\cF = \cF _{31} \ e^3 \wedge e^1 + \cF _{12} \ ( e^1 \wedge e^2 +
 e^3 \wedge e^4 ). \label{cFal}
\ee
In the null Cartesian coordinates the Kerr-Schild null tetrad  has the form
\fn{In the
paper \cite{DKS}
treatment is given in terms of the ``in" - going congruence $e^3$
(advanced fields).
 Here we need to use the ``out''-going congruence. The simplest way to do it
 retaining the basic relations of the paper \cite{DKS} is to replace
$t\to -t$ in the definition of the null Cartesian coordinates.
Therefore, we use here the notations
\bea
2^{1\over2}\z &=& x+iy ,\qquad 2^{1\over2} \Z = x-iy , \nonumber\\
2^{1\over2}u &=& z - t ,\qquad 2^{1\over2}v = z + t . \label{ncc}
\eea
}

\begin{eqnarray}
e^1 &=& d \zeta - Y dv, \qquad  e^2 = d \bar\zeta -  \bar Y dv, \nonumber \\
e^3 &=&du + \bar Y d \zeta + Y d \bar\zeta - Y \bar Y dv, \nonumber\\
e^4 &=&dv + h e^3\label{KSt}.
\end{eqnarray}

Evaluating the basis two-forms in the Cartesian coordinates we obtain
\be
e^1 \wedge e^2 +  e^3 \wedge e^4 = d\z \wedge d \Z + du \wedge dv +
Y d\Z  \wedge  dv,
\ee
and
\be
e^3 \wedge e^1 = Y \ d\Z \wedge d \z + du \wedge dz -
Y du  \wedge  dv - Y^2 \ d\Z \wedge d v.
\ee

\section{Axial stringy system}

The obtained general solution for the aligned electromagnetic fields
(\ref{cFal}) contains the factors which depend on coordinate
$Y = e^{i\phi} \tan \frac {\theta} 2 $ and
can be singular at the z-axis.

We will now be interested in the wave terms and omit the terms describing the
longitudinal components and the field $\gamma$.

The wave terms
are proportional to the following basis  two-forms

$e^3 \wedge e^1 |_{wave} = du \wedge d \z + Y^2 dv \wedge d \Z$

and

$ (e^1 \wedge e^2 + e^3 \wedge e^4) |_{wave} = d \zeta \wedge d \bar\zeta $.

Near the positive half-axis $z^+$,  we have  $Y\to 0$  and
near the negative half-axis $z^-$,  we have  $Y\to \infty$.

Therefore, with the exclusion of the $\gamma$ term, the wave terms
 of the e.m. field (\ref{cFal}) have the form
\be \cF |_{wave} =f_R \ d \z \wedge d u  +
f_L \ d \Z \wedge d v ,
\label{cFLR}
\ee
where the factor
\be
f_R = (AZ),_1
\label{fR}
\ee
describes the ``right"  waves propagating along the $z^+$ half-axis,
and the factor
\be
f_L =2Y \psi (Z/P)^2 + Y^2 (AZ),_1
\label{fL}
\ee
describes the ``left"  waves propagating along the  $z^-$ half-axis,
and some of them are singular at z axis.

Besides, since $Z/P=(r+ia \cos \theta)^{-1}$, all
the terms are also singular at the Kerr ring $r=\cos \theta =0$.
Therefore, the singular excitations of the Kerr ring turn out to be
connected with the axial singular waves.

Let us consider the solutions
describing traveling waves along the Kerr ring
\be
\psi _n (Y,\t) = q Y^n \exp {i\omega _n \t}
\equiv q (\tan \frac \theta 2)^n \exp {i(n\phi + \omega _n \t)}.
\label{psin}
\ee

Near the Kerr ring one has $\psi =\exp {i(n\phi + \omega _n t)}$, and
$|n|$ corresponds to the number of the wave lengths along the Kerr ring.
The parameter $n$ has to be integer for the smooth and single-valued
solutions, however, as we shell see bellow,
the half-integer $n$ can be interesting too.

Meanwhile, by $Y\to 0$ one approaches to the positive z-axis where
the solutions may be singular too.
Similar, by $Y\to \infty$ one approaches to the negative z-axis, and
some of the solutions turns out to be singular there.

When considering asymptotical properties
of these singularities by $r \to \infty $, we have
 $z=r\cos \theta$, and for the distance $\rho$ from the $z^+$ axis we have
the expression $\rho = z \tan \theta \simeq 2 r |Y| $ by $Y\to 0$.
Therefore, for the asymptotical region near the $z^+$ axis we have to put
$Y = e^{i\phi} \tan {\frac \theta 2} \simeq  e^{i\phi} \frac \rho {2r}$, and
$|Y|\to 0$,
while for the asymptotical region near the $z^-$ axis
$Y = e^{i\phi} \tan {\frac \theta 2 } \simeq  e^{i\phi} \frac {2r} \rho  $,
and $|Y|\to \infty$.

 The parameter $\t=t -r -ia \cos \theta$ takes near the
z-axis the values
\be \t _+ = \t |_{z^+}= t-z-ia,\quad \t _- = \t |_{z^-}
=t+z +ia.
\label{tauz}
\ee

It has also to be noted that for $|n|>1$ the solutions contain the axial
singularities which do not fall of  asymptotically, but are increasing.
Therefore, we shell restrict the treatment by the cases $|n|\le 1$.

The leading singular wave for $n=1$ is
\be
\cF^-_1=\frac {4q e^{i2\phi+i\omega _{1} \t_- }} {\rho ^2} \ d \Z \wedge dv.
\ee
This wave propagates to $z=-\infty$ and has the uniform
axial singularity at $z^-$ of order $\rho ^{-2}$.

Meanwhile, the leading singular wave for $n=-1$ is

\be
\cF^+_{-1}= -
\frac {4q e^{-i2\phi+i\omega _{-1} \t_+ }} {\rho ^2} \ d \z \wedge du ,
\ee

and has the similar uniform axial singularity at $z^+$ which
propagates to $z=+\infty$.

The waves with $n=0$ are regular.

In what follows we will show that these singularities form the half-infinite
chiral  strings, in fact superconducting D-strings.
There are several arguments in favor of the system containing a combination
of two strings of opposite chirality, $n=\pm 1$, with
$\omega _1 =-\omega _{-1} = \omega.$

First, if the solution contains only one half-infinite string,
like the Dirac monopole string, it turns out to be asymmetric with respect
to the $z^{\pm}$ half-axis, which leads to a nonstationarity via a recoil.

Then, the symmetric stringy solutions exclude the appearance of monopole
charge.

Similar to the case with the Kerr circular string,
the pure chiral strings, containing modes of only
one direction, cannot exist and any chiral string
has to be connected to some object containing an anti-chiral part.
Indeed, the pure chiral excitation depends only  on one of the
parameters $\t _\pm = t \pm \sigma $, and as a result the
world-sheet is degenerated in a world-line. \fn{This  argument
was suggested by G. Alekseev.} This is seen in the models of the cosmic
chiral strings  where the chiral excitations are joined to
some mass \cite{CarPet} or are sitting on some string having modes
of opposite chirality \cite{BOV}.
In our case the partial pp-wave e.m. excitation has the same
chirality as the half-infinite carrier of this excitation
(the axial ray of PNC).
Therefore, the combination of two $n=\pm 1$ excitations looks very natural
and leads to the appearance of a full stringy system with two
half-infinite singular D-strings of opposite chirality,
``left'' and ``right'', as it is shown at the fig.1.
The world-sheet of the system formed by two straight chiral
strings will be given by
\be
x^\m (t,z)=\frac 12 [ (t-z)k_R^{\m} + (t+z)k_L^{\m}],
\label{astr}
\ee
where the lightlike vectors $k^\m$ are constant and normalized.
At the rest frame the timelike components are equal
$k_R^{0}  =k_L^{0}=1$, and the spacelike components are oppositely directed,
$k_R^{a} + k_L^{a}=0, \quad a=1,2,3.$
Therefore,
$\dot x^\m =(1,0,0,0),$ and $ x'^\m =(0,k^a),$
and the Nambu-Goto string action
\be
S=\alpha^{\prime -1}\int\int\sqrt{(\dot x)^2 ( x')^2 - (\dot x x')^2 } dtdz
\ee
can be expressed via $k_R^{\m}$ and $k_L^{\m}$.

To normalize the infinite string  we have to perform a renormalization
putting      $ \alpha^{\prime -1}\int (x')^2 dz = m, $
which yields the usual action for the center of mass of a pointlike particle
\be
S= m\int \sqrt{( \dot x)^2 }dt.
\ee
As a consequence of this renormalization we obtain that the string tension
$T=\alpha ^{\prime -1} $ is at least close to zero for a free particle.
However, tension can appear in the bounded states where the axial strings
may form the closed loops.

For the system of two straight D-strings  in the rest one can use the gauge
with
$\dot x^0 =1, \quad \dot x^a=0,$ where the term   $(\dot x x')^2$ drops out,
and  the action takes the form
\be
S=\alpha^{\prime -1}\int dt \int \sqrt{p^a p_a} d\sigma,
\ee
where
\be
p^a = \d _\sigma x^a = \frac 12 [ x_R^{\prime\m}(t+\sigma) -
x_L^{\prime\m}(t-\sigma)].
\ee

However, one of the most important arguments in the favor of the above
combination of two chiral strings is suggested by the relation to
the Dirac equation, which must have a physical sense in the structure of the
Kerr spinning particle if it pretends on a classical
description of the electron.

\section{Relation to the Dirac equation}

It is known that in the Weyl basis the Dirac current can be
represented as a sum of two lightlike components of opposite
chirality\fn{We use the spinor notations of the book \cite{WesBeg}.
\be
\gamma_\m  =
\left( \begin{array}{cc}
0 & \sigma _\m \\
\bar\sigma _\m & 0
\end{array} \right) \ ,
\ee
where
\be
\bar\sigma_\m ^{\dot\alpha \alpha} =
\epsilon ^{\dot \alpha \dot \beta} \epsilon ^{ \alpha  \beta}
\sigma _{\m \beta \dot\beta},
\label{bsigma}
\ee
and
\begin{equation}
\begin{array}{cc}
\sigma _0  = \bar\sigma _0 =
\left( \begin{array}{cc}
1 & 0 \\
0 & 1
\end{array} \right) \ ,
\quad
\sigma _1  = -\bar\sigma _1 =
\left( \begin{array}{cc}
0 & 1 \\
1 & 0
\end{array} \right) \ , \\  \nonumber \\
\sigma _2  = - \bar\sigma _2 =
\left( \begin{array}{cc}
0 & -i \\
i & 0
\end{array} \right) \ ,
\quad
\sigma _3  = - \bar\sigma _3 =
\left( \begin{array}{cc}
1 & 0 \\
0 & -1
\end{array} \right) \ .
\label{sigma}
\end{array}
\end{equation}
}

\be
J_\m = e (\bar \Psi \gamma _\m \Psi) = e (\chi ^{+} \sigma _\m  \chi +
\phi ^{+} \bar \sigma ^\m  \phi ),
\ee
where
\begin{equation}
\Psi =
\left(\begin{array}{c}
 \phi _\alpha \\
\chi ^{\dot \alpha}
\end{array} \right),
\label{Psi}\end{equation}
and
\be
\bar\Psi =(\chi ^+,
\phi ^+ )
\label{barPsi}
\ee
It allows one to conjecture that the Dirac equation may describe  the
Kerr axial stringy system - the lightlike currents of two half-infinte
chiral strings. Each of these strings is formed from the spinors which
satisfy the massless Dirac equation. The problem is to get the
four-component spinor which will satisfy the massive Dirac equations
\be(\gamma^\m \hat P_\m -m)\Psi=0,  \qquad
\hat P_\m =i \hbar \d _\m,
\label{Dir}
\ee
which split in this basic into two systems
\be
m \phi _\alpha  =
i \sigma ^\m _{\alpha \dot \alpha} \d_\m \chi ^{\dot \alpha}, \qquad
m \chi ^{\dot \alpha}  =
i \bar\sigma ^{\m \dot\alpha \alpha}  \d_\m \phi _{\alpha},
\label{DirSpl}
\ee
Let us recall now
that the used in the Kerr-Schild null tetrad function $Y$ (\ref{KSt})
is the projective spinor coordinate
$Y=\phi _2 /\phi _1$
\cite{Bur-nst}.
Near the $z^+$ half-axis we have $Y\to 0$, and one can set in this limit
\begin{equation}
{\phi}_{\alpha} =
\left(\begin{array}{c}
 {\phi}_{1}\\
\phi _{2}
\end{array} \right) =
\left(\begin{array}{c}
 1\\
0
\end{array} \right),
\qquad
{\phi}^{\alpha} =
\epsilon ^{\alpha \beta}{\phi}_{\beta} =
\left(\begin{array}{c}
 0\\
-1
\end{array} \right) .
\label{spYR}
\end{equation}
This spinor describes the lightlike vector
\be
k_{R} =d(t-z)=
\bar  \phi _{\dot\alpha}
\bar\sigma _\m^{\dot\alpha\alpha} dx^\m  \phi _\alpha =
(1,0) \bar\sigma _\m dx^\m
\left(\begin{array}{c}
 1\\
0
\end{array} \right),
\label{kR}
\ee
since
\be
\sigma _\m dx^\m =
\left(\begin{array}{cc}
dt+dz & dx-idy\\
dx+idy & dt-dz
\end{array} \right), \qquad
\bar\sigma _\m dx^\m =
\left(\begin{array}{cc}
dt-dz & -dx+idy\\
-dx-idy & dt+dz
\end{array} \right).
\label{sigdx}
\ee
Similar, near  the $z^-$ half-axis we have $Y\to \infty$,
and this limit corresponds to the spinor
\begin{equation}
{\chi}_{\alpha} =
\left(\begin{array}{c}
 0\\
1
\end{array} \right),
\qquad
{\chi}^{\alpha} =
\left(\begin{array}{c}
 1\\
0
\end{array} \right) ,
\label{spYL}
\end{equation}
which describes the lightlike vector\fn{Note, that for commuting spinors
$\chi \sigma_\m \bar \phi =\bar \phi \bar\sigma_\m \chi .$}
\be
k_{L} =d(t+z)=
\chi ^{\alpha}
\sigma _{\m\alpha\dot\alpha} dx^\m  \bar\chi ^{\dot\alpha} =
(1,0) \sigma _\m dx^\m
\left(\begin{array}{c}
 1\\
0
\end{array} \right).
\label{kL}
\ee
The lightlike vectors $k_L$ and $k_R$ are the generators of the
left and right chiral half-infinite strings correspondingly.

Since the spinor functions $\psi$ and $\chi$ are fixed up to arbitrary
gauge factors, one can form the four component spinor function
\begin{equation}
\Psi = {\M}(p_\m x^\m)
\left(\begin{array}{c}
a \phi _\alpha \\
b \chi ^{\dot \alpha}
\end{array} \right),
\label{MPsi}\end{equation}
which may satisfy the Dirac equation.

Indeed, substitution  (\ref{spYR}), (\ref{spYL}) and  (\ref{MPsi}) into
(\ref{DirSpl}) leads to the equations
\be
am=(p_0 -p_z) b \ln {\M}' , \qquad
bm=(p_0 +p_z) a \ln {\M}', \qquad
p_x +ip_y =0 ,
\label{Dspl}
\ee
which realize the Dirac idea on splitting the
relation   $p_0^2 = m^2 +p_z^2. $
Other necessary conditions yield
$p_x =p_y =0. $  One sees that this solution corresponds to an
arbitrary relativistic motion along the axial stringy system, which retains
axial symmetry of this system.
The function ${\M}=e^{-i \omega t + i z p_z }$  oscillates with the
Compton frequency which is determined by excitations of the Kerr
circular string and contains a
a plane fronted modulation of the chiral strings by de Broglie periodicity.

One can conjecture which changes could be performed to get the
 Kerr anti-particle. It has to be the change of the PNC direction
$k_{out} \to k_{in}$, which can be achieved by the transition to the
negative sheet of the metric. It yields a natural picture of annihilation
as it is shown in the fig.3. It was discussed in \cite{BurMag} that
the size of the Kerr circular string for the massless Kerr spinning
particle has to grow to infinity and disappear. As a result there retains
only a single chiral string which may correspond to a massless particle.

\begin{figure}[ht]
\centerline{\epsfig{figure=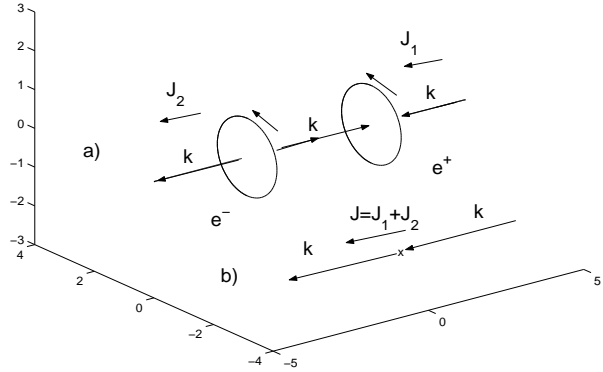,height=6cm,width=8cm}}
\caption{ (a) annihilation of the Kerr particle and antiparticle and
(b) formation of the lightlike particle.}
\end{figure}

To form a symmetric solution we used a combination of the $n=\pm 1$
excitations

\be \psi _n (Y,\t)(Z/P)^2
\simeq q e^{i n\phi + i\omega _n \t} 2^{-n}\rho ^n /r^{n +2}
\ee
with $\omega _1 =-\omega _{-1} = \omega.$
Indeed, also the solution
containing combination of three terms with $n=-1, 0, 1$ and $\omega _0 =0$
represents especial interest since it yields
a smooth e.m. field packed along the Kerr string with
one half of the wavelength and gives an electric charge to the solution.

Note, that orientifold structure of the Kerr circular string
admits apparently the excitations with $n= \pm 1/2$ too, so far as the
negative half-wave can be packed on the covering space turning into
positive one on the second sheet of the orientifold. However, the meaning
of this case is unclear yet, and it demands a special consideration.

\section{Einstein-Maxwell axial pp-wave solutions}

The e.m. field given by (\ref{cFLR}), (\ref{fR}) and (\ref{fL}) can be
obtained from the potential

\be
\cA= - AZ e^3 - \chi d \Y ,
\label{cA}
\ee
where $A=\psi /P^2$ is given by (\ref{A}) and
\be
\chi = \int P^{-2} \psi d Y,
\label{chi}
\ee
$\Y$ being kept constant in this integration.
The considered wave excitations have the origin from the term
\be
\cA= P^{-2}\psi _n Z e^3 = q Y^n \exp {i\omega \t} P^{-2} Z e^3
\ee
and acquire the following asymptotical $z^{\pm}$ forms:

For $n=1; z<0$

\be
\cA ^- =q Y e^{i\omega _n \t}(r+ia \cos \theta)^{-1} e^3/P \simeq
-2 q \frac { e^{i\omega _1 \t_- +i\phi}}{\rho} dv .
\ee

For $n=-1; z>0$

\be
\cA ^+ =q Y^{-1} e^{i\omega _n \t}(r+ia \cos \theta)^{-1} e^3/P \simeq
2 q \frac { e^{i\omega _{-1}\t_+ -i\phi}}{\rho} du.
\label{Au}
\ee

Each of the partial solutions
represents the singular plane-fronted e.m. wave propagating along
$z^+$ or $z^-$ half-axis without damping.
It is easy to point out the corresponding
self-consistent solution of the Einstein-Maxwell field equations which
belongs to the well known class of pp-waves \cite{KraSte,Per}.

The metric has the Kerr-Schild form
\be g_{\m\n}
= \h_{\m\n} + 2 h k_{\m} k_{\n}, \label{ks} \ee
where function $h$ determines the Ricci tensor
\be
R^\mn = - k^\m k^\n \Box h,
\label{Rmn}
\ee
 $k^\m= e^{3\m}/P$ is the normalized principal null direction
(in particular, for the $z^+$ axis $k^\m dx^\m= - 2^{1/2}du $), and
$\Box$  is a flat D'Alembertian
\be
\Box=2\d _\z \d _\Z + 2 \d _u \d _v  \ .
\label{Box}
\ee
The Maxwell equations take the form
\be
\Box \cA = J=0
\label{BoxA}
\ee
and can easily be integrated leading to the solutions
\bea
 \cA ^+ = [ \Phi ^+(\z) + \Phi ^-(\Z) ] f^+(u,v) du, \\
 \cA ^- = [ \Phi ^+(\z) + \Phi ^-(\Z) ] f^-(u,v) dv,
\label{cApm}
\eea
where $\Phi ^{\pm}$ are arbitrary analytic functions, and functions
$f^\pm $
describe the arbitrary retarded and advanced waves.
In our case we have the retarded-time parameter
$\t = t - r - ia \cos \theta$ which takes at the $z^+$
axis the values $\t \simeq - 2^{1/2} u -ia $ and at the $z^-$
axis the values $\t \simeq 2^{1/2} v + ia $. Therefore, we have
\be
f^+ = f^+(u), \quad f^- = f^-(v).
\label{fpm}
\ee
The corresponding energy-momentum tensor  will be
\be
T^\mn = \frac 1 {8\pi} |\cF^+_{-1}|^2 k^\m k^n ,
\label{Tmn}
\ee
where for $z^+$ wave $k_\m dx^\m = -2^{1/2}du $ and
for $z^-$ wave $k_\m dx^\m = 2^{1/2}dv $.

The Einstein equations $R^\mn = -8 \pi T^\mn$ take the simple
asymptotic form
\be
\Box h = |\cF^+_{-1}|^2 = 16 q^2e^{-2a\omega}\rho^{-4}.
\label{heq}
\ee
This equation can  easily be integrated and yields the singular solution
\be h=  8 q^2e^{-2a\omega}\rho^{-2} .
\label{h}
\ee
Therefore, the wave excitations of the Kerr ring lead to the appearance
of singular pp-waves which propagate outward along the $z^+$ and/or
$z^-$ half-axis.

These axial singularities are evidences
of the axial stringy currents, which are exhibited explicitly
when we  regularize the singularities on the base of the
Witten field model for the cosmic superconducting strings \cite{Wit,VS}.

The resulting excitations have the Compton wave length which is
determined by the size of the Kerr circular string.
However, for the moving systems the excitations of the axial stringy system
are modulated by de Broglie periodicity.

\section{Regularization. Superconducting strings as sources of axial pp-waves}

The singular lines are  often considered as strings.
It is assumed that singularity
is only an approximation which has to be replaced by a regular matter source
in a more general field model. This point of view was
considered for interpretation of the Dirac monopole singular line, and
the other very well known example is the Nielsen-Olesen string
model in the form of a vortex line in superconductor \cite{NilOle}.
Both these examples are related to our case and the Higgs field model for
the stringy sources is used as the most simple and adequate to these cases.
Many physical stringy models as well as models of bags and domain walls
are based on the different modifications of the Higgs field model.
Almost all of the models leads to the picture of an extended physical object
(string,  bag or other) which is suited in the vacuum possessing
superconducting properties. Therefore, the infinite external space surrounding
these objects turns out to be superconducting.
Although there is no usually the exact analytical
solutions, the models turn out to be rather simple and solvable numerically.
Meanwhile, the main physical discrepancy is not discussed usually: the
real surrounding vacuum is not superconducting! Electromagnetic fields are
freely propagating there, they are long range and do not acquire a mass from
the Higgs field as it follows from these models. The physically right
picture ought to be turned over: the superconducting object is to be
surrounded by a real vacuum with long range electromagnetic field.
Therefore, the considered usually models of extended objects describe the
picture which is ``dual'' in some sense to the real physical situation\fn{We
are sorry for use the word ``dual'' which has to many other physical
meanings.}  Meanwhile, in spite of this seeming failure, the use of
these dual models is striking productive. The resolution of this
contradiction lies in the assumption that there are two type of
superconductivity, the usual ``true'' one and ``dual" or ``false'' one,
and our observable ``true'' vacuum is not superconducting with respect
to the usual electromagnetic field, $U(1)$ gauge, but possesses a
``dual'' or ``false'' superconductivity with respect to some other
$\tilde U(1)$ gauge field which can be confined on the extended objects.

A model of this type,
the $U(1)\times \tilde U(1)$ field model,
 was suggested by Witten for the cosmic
superconducting strings \cite{Wit,VS} and
represents a doubling of the usual Abelian  Higgs model.

It  contains
two sectors, say $A$ and $B$, with two Higgs fields $\phi_A$ and $\phi_B$,
and two gauge fields $A_{\mu}$ and $B_{\mu}$ yielding two sorts
of superconductivity $A$ and $B$.
The gauge field $A_{\mu}$ of the $A$ sector is the usual
electromagnetic field which is long range in outer region and acquires
a mass interacting with the chiral scalar field of this sector
$\phi_A$. This scalar field $\phi_A$ can be concentrated on the extended
superconducting objects and describes the superconducting vacuum state.
\par
The sector $B$ of the model describes a ``dual'' picture.
The nonzero chiral field $\phi_B$ covers almost all our space for the
exclusion of the local regions occupied by superconducting objects, or
for the exclusion of the regions of localization of the field $\phi_A$.
Therefore, the localizations of the vacuums $A$ and $B$
are dual to each other. Similarity of the sectors $A$ and $B$ allows one
to consider field $\phi _B$ also as a carrier of some kind of
superconductivity, but this is a ``false'' or ``dual'' superconductivity
which covers almost all our empty space. Therefore, our usual physical vacuum
is considered as a ``dual'' superconductor in this model.

\par
The corresponding Lagrangian of the Witten $U(I)\times \tilde U(I)$
field model is given by
\cite{Wit}
\begin{equation}
L=-(D^\mu \phi_A )(\overline{ D_\mu \phi_A} )-(\tilde D^\mu \phi_B )
(\overline{\tilde D_\mu \phi_B} )-\frac 14 F_A^{\mu \nu }F_{A\mu \nu }-
\frac 14 F_B^{\mu \nu } F_{B\mu \nu}-V,
\label{WL}
\end{equation}
where
$F_{A\mu \nu }=\partial _\mu A_\nu -\partial _\nu A_\mu $ and
$F_{B\mu \nu}=\partial _\mu B_\nu -\partial _\nu B_\mu $
are field stress tensors, and the potential has the form
\begin{equation}
V=\lambda (\bar{\phi_B}\phi_B -\eta ^2)^2+f(\bar{\phi_B}
\phi_B -\eta ^2)\bar{\phi_A}\phi_A +m^2\bar{\phi_A}\phi_A +\mu
(\bar{\phi_A}\phi_A )^2.  \label{V}
\end{equation}
Two Abelian gauge fields $A_\mu$ and $B_\mu$ interact separately
with two complex scalar fields $\phi _B$ and $\phi_A$ so that the
covariant derivative $D_\mu \phi_A =(\partial +ie A_\mu) \phi_A $ is
associated with $A$ sector, and covariant derivative
$\tilde D_\mu \phi _B = ( \partial +ig B_\mu) \phi _B $ is associated with
$B$ sector.
Field $\phi _B $ carries a $\tilde U(1)$ charge $\tilde q \ne 0$ and a
$U(1)$ charge $q=0$,
 and field $\phi_A $ carries $\tilde U(1)$ and $U(1)$ charges of
$\tilde q=0$ and $q\ne 0$, respectively.
\par

The A and B sectors are almost independent interacting only
through the potential term for scalar fields. This interaction has to
provide a synchronized phase transitions from superconducting B-phase
inside the bag to superconducting A-phase in the outer region.
The synchronization of this transition occurs explicitly in a supersymmetric
version of this model given by Morris \cite{Mor}.  Application
of this model to the Kerr source is discussed in \cite{BurBag}.\fn{
As it showed the treatment in  \cite{BurBag},  the both (false and true) vacua
are to be the supersymmetric states. These vacua are separated by a
D2-brane (domain wall) which has to take a tubelike form by application
of this field model to the regularized axial pp-waves. The resulting
structure turns out to be similar to the supersymmetric tubes considered in
\cite{MatTow}.}

In this section we consider in details only $A$ sector
which describes the singular  e.m. excitations of the Kerr ring
interacting with a superconducting stringlike source, which regularizes
these singularities. For simplicity we restrict ourself by the treatment
in a flat background.
We put $\phi _A=\Phi e^{i\chi}$ and ignore the
gravity and the fields of sector $B$. The
necessary form of the potential $V$ we will discuss later.

The field equations in this case take the form
\be
D_\m D^\m \Phi e^{i\chi} + \frac 12 \partial _\phi V =0,
\label{Dphi}
\ee
and
\be
\Box A_\m  = J_\m = e \Phi ^2 (-2 \chi ,_\m + e A_\m).
\label{dA}
\ee
It is known that the problems of this sort do not have explicit analytical
solutions \cite{BOV}.
We consider the far zone near the $z^+$-axis, and from
(\ref{Au}) we have the only singular nonzero
component of the massless e.m. field  in the far zone
$A_u = \frac {2 q} \rho e^{i\omega \t_+ -i\phi}$.
By using (\ref{tauz}) and the coordinates $\z=2^{-1/2}\rho e^{i\phi}$
and $\Z=2^{-1/2}\rho e^{-i\phi}$ we obtain
\be
A_u = A_u^{(0+)} = \frac {2C} \z e^{-i2^{1/2}\omega u},
\label{Au+}
\ee
where
\be
C= 2^{-1/2} q e^{a\omega}.
\label{Cu}
\ee
The real vector potential will be
\be
A_u^{(0)} =\frac 12 (A_u^{(0+)}+A_u^{(0-)}),
\label{rAu}
\ee
where
\be
A_u^{(0-)}=\frac {2C} \Z e^{i2^{1/2}\omega u}.
\label{Au-}
\ee
Computations turn out to be easy in this form.
Index $^{(0)}$ we use to underline that this component is long range and
massless.
The physics of the process suggests us the way to solve these equations.
The singular wave $A_u^{(0)}$ is to be regularized penetrating into
the Higgs field $\Phi e^{i\chi}$ which is condensed on the string.
Regularization can be performed by a mechanism of compensation which is
similar to the Feynman and Pauli-Villars scheme of regularization.
A massive short range field  $A_u^{(m)}$, having just the same structure
near the singularity as the field $A_u^{(0)}$, has to compensate
singularity leading to a regular behavior of the sum $A_u^{(0)}+A_u^{(m)}$.

Therefore, we have to split the eq.(\ref{dA}) in the sum of two equations:

- one of them for the massless components $A_u^{(0\pm)}$,
\be
\Box A_u^{(0)}  = 0 = e \Phi ^2 (-2\chi ,_u + e A_u^{(0)}),
\label{dA0}
\ee
and another one for the massive field
\be
\Box A_\m^{(m)}  = J_{\m}= e \Phi ^2 (- 2\chi ,_\m + e A_\m^{(m)}).
\label{dAm}
\ee
From the first equation we get $2 \chi,_u = A_u^{(0)}$, which can be easily
integrated leading to
\be
\chi = \chi_0 (\z,\Z) + \frac e2 \int A^{(0)} du = \chi_0 (\z,\Z) +
\frac e {2i\omega} [A_u^{(0+)} - A_u^{(0-)}].
\label{chi1}
\ee
Since the term $\chi,_u$ has been already eaten by the term
$A_u^{(0)}$ we have got for the massive u-component the equations
\be
\Box A_u^{(m\pm)}  = J_u^{\pm}= e^2 \Phi ^2  A_u^{(m\pm)}.
\label{dAmu}
\ee
However, the mass term   $ e^2 \Phi ^2 $ is not constant in our case, but
is a function of $\rho$ which fall off outside the string core.
It complicates the regularization procedure since
the solution of this equation will represent an interpolation
between some solution with mass at the string core and the massless solution
in far zone.

To solve this equation we use the ansatz
\be
A_u^{(m\pm)} =e^{f_\pm (\rho)} A_u^{(0\pm)}
\label{Aumpm}
\ee
which will provide a similarity of the massive and massless components near
the string core if the function $f_\pm (\rho)$ tends to zero by $\rho \to 0$.

Therefore, from (\ref{dAmu}) we obtain the equations
\be
\Box e^{f_\pm} A_u^{(0\pm)}  = e \Phi ^2 e^{f_\pm} A_u^{(0\pm)}.
\label{dfA0}
\ee
It turns out that we can use a common function $f(\rho)$, or
$f_+ = f_- =f(\rho^2)\equiv f(\z\Z/2)$.\fn{By using the coordinates $\z,\Z$
and  the relations
$\z\Z=2\rho^2,\quad \rho,_\z=\frac\Z {4\rho}, \quad
\rho,_{\z\Z} =\frac 1{8\rho}$ computations are simplified.}

\begin{figure}[ht]
\centerline{\epsfig{figure=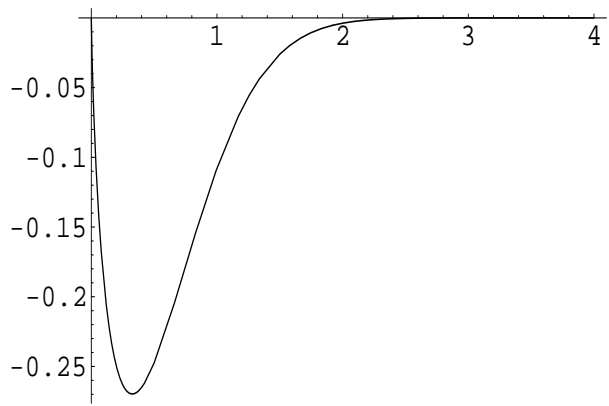,height=6cm,width=8cm}}
\caption{Regularizing dependence $f'(\rho)$ for  the Gaussian
distribution of condensate on the string.}
\end{figure}

Taking into account (\ref{dA0}) we obtain from (\ref{dfA0}) the following
differential equation
\be
f'' + (f')^2 - f'/\rho =8 e^2 \Phi^2 .
\label{ddf}
\ee
Assuming for the shape of superconducting
condensate on the string the distribution
\be
\Phi(\rho)=c e^{-\rho^2},
\label{Phi}
\ee
we obtain by numeric computations the following shape of the
function  $f'(\rho)$, which is shown on the fig.4.

The resulting regularized u-component takes the form
\be
A_u =(1 -e^{f (\rho)}) A_u^{(0)},
\label{Autot}
\ee
 and singularity cancels.

As it is seen from the fig.4, in the far zone the function 
$f(\rho)= \int _0 ^\rho f^\prime(x)dx$ will tend to some negative constant, 
and as the result the factor $ \ 1-e^{f(\rho)} $ will
partially suppress the solution outside the string core too.

\begin{figure}[ht]
\centerline{\epsfig{figure=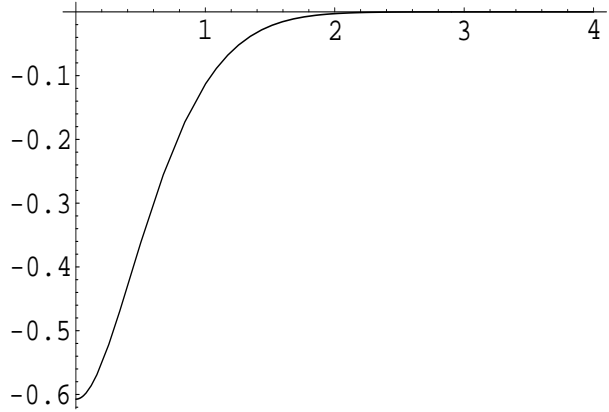,height=6cm,width=8cm}}
\caption{The dependence of massive components
$A_\z ^{(m)}$ and $A_\Z ^{(m)}$ on the axial distance $\rho$.}
\end{figure}

For the other massive components the equations will retain the form
(\ref{dAm}), and
from   (\ref{chi1}) one obtains
\be
\chi,_\z = \chi_0 (\z,\Z),_\z -
\frac e {2i\omega \ \z} A_u^{(0+)}, \qquad
\chi,_\Z = \chi_0 (\z,\Z),_\z +
\frac e {2i\omega \ \Z} A_u^{(0-)}.
\label{dchi}
\ee
Numerical computations allow us to obtain  the
behavior of the massive components $A_\z ^{(m)}$ and $A_\Z ^{(m)}$.
The result of regularization is shown on the fig.5.
One sees that the used ``compensatory'' approach is really effective,
and one can expect that it will also be effective for regularization of the
Kerr ringlike singularity and by the formation of the Kerr's
disklike source.

\section{Conclusion}

The considered axial stringy system of the Kerr spinning particle
gives a new view on the physical sense of the Dirac equation, and also
maybe on the physical sense of the quantum wave function,
divergences and  regularization.

It should be also noted a striking similarity  of this structure
with the well known
elements and methods of the signal transmission in the systems of radio
engineering and radar systems. In fact, the chiral axial string
resembles a typical system for the signal transmission containing
a carrier frequency which is modulated by a signal - the wave function
which is the carrier of information. The Kerr circular string also plays
the role of a generator of the carrier frequency.

Basing on the principle that
the fine description of a quantum system has to absorb  maximally the
known classical information on this system, one can conjecture
that the above strikingly simple structure may have a
relation to the structure of spinning particles.
In this scheme the quantum wave function has a physical carrier.
It should be mentioned that the considered
topological coupling of the circular Kerr string and the axial
stringy system reproduces the very old de Broglie's ``wave-pilot''idea on
a singular  carrier of the wave function which
controls the dynamics of quantum particle.

We have already mentioned, that the axial string tension tends to zero
for a free
particle, but it has to be finite for the bounded states when the axial
string can form the closed
loops . One can also expect that for the closed loops
an extra tension can
appear caused by the magnetic flows which can concentrate on these strings
\cite{GarVac}, and it has to be a question for further investigations.
The role of  the axial strings in the mass renormalization, as well as the
gravitational (pp-wave) nature of the axial strings contain a hint that such
closed loops of the strings may be candidates for gravitons.
One can also assume that vacuum itself may be formed from the
strings of this type.

The existence of the axial stringy system in the Kerr spinning particle
allows us to overcome the old contradiction between the
 very small experimental value  of the electron cross section and the
very large Compton size of the Kerr source.
One can conjecture now that the cross section
of electron may be determined by the contact stringy
interaction of the axial stringy system, acquiring also some corrections
from the Compton region of the Kerr circular string.

It seems that the axial stringy system
may be observable in the experiments with scattering of the soft laser beams
on the crossing beams
of polarized electrons and maybe on the crossing laser beams too, and this
may be crucial for verification the reality
of this axial stringy system.

\section*{Acknowledgments}
This work was  supported from the
Internet Science Education Project (ISEP) and we are very thankful to
Jack Sarfatti for this support and useful discussions. The work was also
reported on the seminar of the Russian Gravitational Society (the firmer
seminar by Prof. Dmitrii Ivanenko), and author is thankful to participants
of this seminar for very active discussion. We are also thankful to
G. Alekseev, B.N. Frolov and S. Odintsov for useful conversations, and
especially to S.R. Hildebrandt for the thorough reading the paper and
worthwhile remarks. We would also like to thank O. Lunin for very
stimulating communications leading to the better understanding
the relation of this model to the fife-dimensional rotating solutions.


\begin{thebibliography}{99}
\bibitem{Car}
B. Carter, Phys. Rev. {\bf 174} (1968) 1559.
\bibitem{Isr}
W. Israel, Phys. Rev. {\bf D2} (1970) 641.
\bibitem{Bur0}
   A.Ya. Burinskii, Sov. Phys. JETP,  {\bf39}(1974)193.
\bibitem{IvBur} D. Ivanenko and A.Ya. Burinskii,
   Izvestiya Vuzov Fiz. n.5 (1975) 135 (in russian).
\bibitem{Lop}
C.A. L\'opez, Phys. Rev. {\bf D30} (1984) 313.
\bibitem{BurSen}
 A. Burinskii, Phys.Rev.{\bf D 52} (1995)5826.
\bibitem{BurStr}
A.Ya. Burinskii,
Phys.Lett. {\bf A 185} (1994) 441;
{\it String-like Structures in Complex Kerr Geometry.}
In: ``Relativity Today'',  Edited by R.P.Kerr and Z.Perj\'es,
Akad\'emiai Kiad\'o, Budapest, 1994, p.149.
\bibitem{BurSup}
 A. Burinskii,
Phys.Rev.{\bf D 57} (1998)2392;
 Class.Quant.Grav. {\bf 16}(1999)3497.
\bibitem{BurBag}
A. Burinskii, Grav.\& Cosmology.{\bf 8} (2002) 261.
\bibitem{IvBur1} D.
   Ivanenko and A.Ya. Burinskii, Izvestiya Vuzov Fiz. n.7 (1978) 113 (in
   russian).
\bibitem{Bur1}  A.Ya. Burinskii, {\it Strings in the
Kerr-Schild metrics} In: ``Problems of theory of gravitation and elementary
   particles'',{\bf11}(1980), Moscow, Atomizdat, (in russian).
\bibitem{BurOri}
 A. Burinskii,
Phys.Rev.{\bf D 68} (2003)105004.
\bibitem{BurTwo} A. Burinskii, {\it Two Stringy Systems of the Kerr Spinning
Particle}, submitted to the Proc. of the XXVI Workshop on Fundamental
Problems of HEP and Field Theory, IHEP, Protvino 2003, hep-th/0402114.
\bibitem{DKS}  G.C. Debney, R.P. Kerr, A.Schild, J. Math.
Phys. {\bf10}(1969) 1842.
\bibitem{Bur-nst}
A. Burinskii, Clas.Quant.Gravity
{\bf 20} (2003)905;
Phys. Rev.
{\bf D 67} (2003) 124024.
\bibitem{KraSte} D.Kramer, H.Stephani, E. Herlt, M.MacCallum, ``Exact
Solutions of Einstein's Field Equations'', Cambridge Univ. Press,  1980.
\bibitem{Per} A. Peres, Phys. Rev. {\bf 118} (1960)1106.
\bibitem{CarPet}B.Carter, P.Peter,
Phys. Rev. {\bf D 52}(1995)1744.
\bibitem{Vil} A. Vilenkin, Nucl. Phys. {\bf
B249}(1985)
\bibitem{BOV} J.J. Blanco-Pillado, Ken D. Olum, Alexander Vilenkin
Phys. Rev. {\bf D 63}(2001)103513.
\bibitem{Wit}
E. Witten, Nucl.Phys.,{\bf B249}(1985)557.
\bibitem{VS}
A. Vilenkin and E.P.S. Shellard, {\it Cosmic Strings and Other Topological
Defects} (Cambrige University Press, 1994) hep-ph/9706302
\bibitem{Dab} A. Dabholkar, J.P Gauntlett, J.A. Harvey and Waldram,
{\it Strings as Solitons and Black Holes as Strings}, Nucl. Phys. {\bf B 474}
(1996)85, hep-th/9511053.
\bibitem{HorTse} G.T.Horowitz and A.A.Tseytlin
Phys. Rev. {\bf D 51} (1995)2896, hep-th/9409021.
\bibitem{LunMat}O.Lunin and S.D.Mathur,
Nucl. Phys. {\bf B 610}(2001)49, hep-th/0105136.
\bibitem{MatTow} D. Mateos, S. Ng and P. K. Townsend, {\it Supercurves}
Nucl. Phys. {\bf B 610}, hep-th/0204062
\bibitem{Sch} S.S. Schweber, ``An Introduction to Relativistic Quantum Field
Theory", Row, Peterson and Co.Evanston, Ill.,Elmsford, NY, 1961.
\bibitem{AhiBer} A.I. Akhiezer and V.B. Berestetsky, Quantum Electrodynamics, Moscow
"Nauka", 1981 (in Russian).
\bibitem{Mor}
J.R. Morris, Phys.Rev.{\bf D 53}(1996)2078, hep-ph/9511293;
\bibitem{Sen}
A. Sen,  Phys. Rev. Lett. {\bf 69} (1992) 1006.
\bibitem{GSW} M.B. Green, J.h. Schwarz, and E. Witten, `Superstring
Theory', V. I, II,  Cambridge University Press, 1987.
\bibitem{BurMag} A. Burinskii and G. Magli,
Phys. Rev. {\bf D 61}(2000)044017.
\bibitem{NilOle}
H.B.Nielsen and P. Olesen, Nucl. Phys.,{\bf B61}(1973)45.
\bibitem{WesBeg}
J. Wess and J.Bagger, Supersymmetry and Supergravity, Princeton,
New Jersey 1983.
\bibitem{GarVac} J. Garriga and T. Vachaspati, Nucl.Phys.,{\bf B438}(1995)161,
hep-ph/9411375.
\end{thebibliography}
\end{document}